\documentclass[
    ,final            % use final for the camera ready runs
  ]
  {aipproc}

\layoutstyle{6x9}

\def\la{\mathrel{\hbox{\rlap{\hbox{\lower4pt\hbox{$\sim$}}}\hbox{$<$}}}}

\begin{document}

\title{The UCT Survey of Old Novae}

\author{Patrick A. Woudt}{
  address={Department of Astronomy, University of Cape Town, Rondebosch 7700, 
South Africa}
}
\author{Brian Warner}{
  address={Department of Astronomy, University of Cape Town, Rondebosch 7700, 
South Africa}
}

\begin{abstract}
We present a status report on our high speed photometric survey of faint
Cataclysmic Variables, which is concentrating on old novae.
\end{abstract}

\maketitle

The high speed photometric survey of southern Cataclysmic Variable stars, 
of which one paper has been published (Woudt \& Warner 2001) and another is 
in press (Woudt \& Warner 2002), uses the UCT CCD photometer attached to the 
1.0-m and 1.9-m reflectors at the Sutherland site of the South African 
Astronomical Observatory. We have concentrated on faint old novae in 
crowded fields; these have generally been neglected in the past as being 
difficult to observe, but they prove to be a rich source of 
interesting phenomena, especially relating to orbital modulations 
and magnetic effects.
     
\begin{figure}[h]
  \includegraphics[height=.3\textheight]{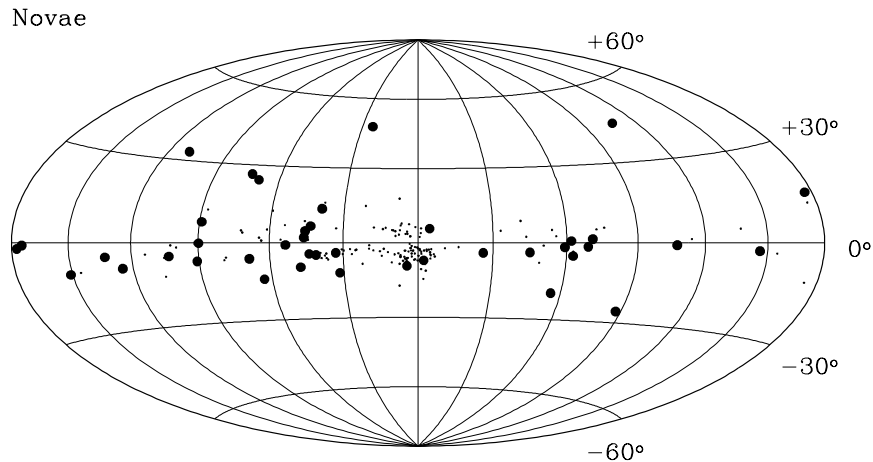}
  \caption{The distribution of old novae in Galactic coordinates (Aitoff projection).
The Galactic Centre is at the centre of the figure, with increasing longitudes to the
left. Novae with known orbital periods are indicated by the large filled circles.}
  \label{fig2}
\end{figure}

From photometric modulations we have determined orbital periods for 
the old novae RS Car (N1895), V365 Car (N1948), RR Cha (N1953), 
BY Cir (N1995), DD Cir (N1999), AP Cru (N1936), CP Cru (N1996), 
V351 Pup (N1991), V630 Sgr (N1936), V697 Sco (N1941) and V992 Sco (N1992). 
Details of these can be found in Table 1 of the Review paper by Warner 
in these Proceedings.
The Galactic distribution of old novae is shown in Fig.~\ref{fig2}, where those 
with known orbital periods are shown as large filled circles. 
The southern sky is no longer 
so undersampled as it was a few years ago.

\begin{figure}
  \includegraphics[height=.3\textheight]{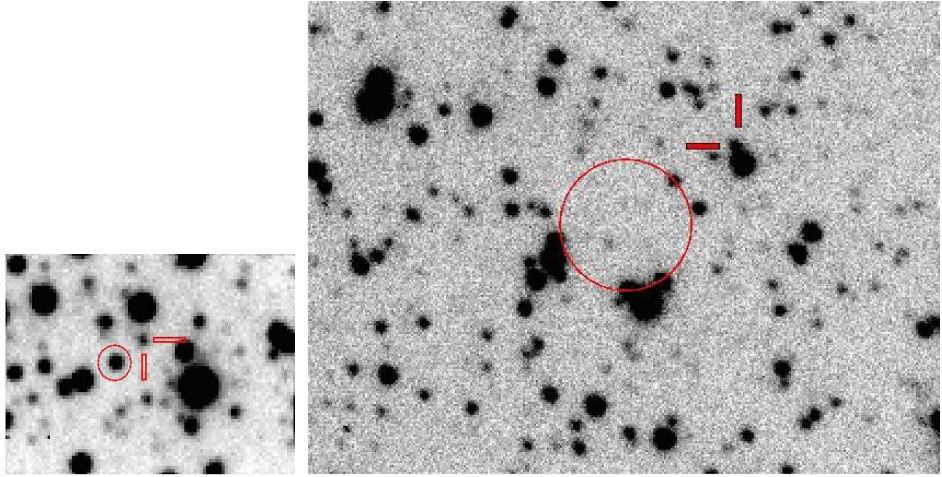}
  \caption{Finding charts of MT Cen (left panel)
and X Cir (right panel). MT Cen was observed with the 1.9-m telescope 
(f.o.v. $50'' \times 34''$), X Cir with the 1.0-m telescope (f.o.v. 
$109'' \times 74''$). Previous positions are indicated by circles, updated
indentifications are indicated by markers. }
  \label{fig1}
\end{figure}

In addition, from high speed flickering activity we have been able to 
determine the correct identification for 4 previously misidentified or 
unidentified old novae; MT Cen, X Cir, V552 Sgr and CQ Vel. Finding charts
for V552 Sgr and CQ Vel are shown in Woudt \& Warner (2001; 2002, respectively).
Finding charts for MT Cen and X Cir are shown in Fig.~\ref{fig1}.
     
Apart from the new orbital periods, we have also found periodicities 
that are ascribed to the rotation periods of the white dwarf components. 
These are the signatures of intermediate polars, which have magnetic 
primaries. Table 1 gives details of the spin periods found during our survey.

\begin{table}
\begin{tabular}{lcrlc}
\hline
\tablehead{1}{l}{b}{Star} &
\tablehead{1}{c}{b}{$P_{spin}$ (sec)} &
\tablehead{1}{r}{b}{\ \ \ } &
\tablehead{1}{l}{b}{Star} &
\tablehead{1}{c}{b}{$P_{spin}$ (sec)} \\
\hline
RX\,J1039.7-0507     &    1444     &    &  RR Cha & 1950 \\
AP Cru               &    1850     &    &  V697 Sco & $\sim$6000   \\
\hline
\end{tabular}
\caption{Old novae with newly detected spin periods.}
\label{tab1}
\end{table}

\begin{theacknowledgments}
PAW is funded by the National Research Foundation and by strategic funds made available to BW. BW is funded entirely by the University of Cape Town.
\end{theacknowledgments}

\bibliographystyle{aipprocl} % if natbib is missing

\end{document}